\documentclass[aps,prb,twocolumn,superscriptaddress]{revtex4}
\usepackage{amsmath}
\usepackage{dcolumn}
\usepackage{epsfig}
\usepackage{graphicx}
\usepackage{latexsym}

\begin{document}

\title[Focus on Superconductors with Exotic Symmetries]{Polar Kerr Effect as Probe for Time-Reversal Symmetry Breaking in Unconventional Superconductors}

\author{Aharon Kapitulnik}
\affiliation{Department of Applied Physics, Stanford University, Stanford, CA 94305}
\affiliation{Department of Physics, Stanford University, Stanford, CA 94305}
\author{Jing Xia}
\affiliation{Department of Physics, Stanford University, Stanford, CA 94305}
\author{ Elizabeth Schemm}
\affiliation{Department of Physics, Stanford University, Stanford, CA 94305}
\author{Alexander Palevski}
\affiliation{School of Physics and Astronomy, Tel Aviv University, Tel Aviv 69978, Israel}

\begin{abstract}
The search for broken time reversal symmetry (TRSB)  in unconventional superconductors intensified in the past year as more systems have been predicted to possess such a state. Following our pioneering study of TRSB states in Sr$_2$RuO$_4$ using magneto-optic probes, we embarked on a systematic study of several other of these candidate systems. The primary instrument for our studies is the Sagnac magneto-optic interferometer, which we recently developed. This instrument can measure magneto-optic Faraday or Kerr effects with an unprecedented sensitivity of 10 nanoradians at temperatures as low as 100 mK.  In this paper we  review our recent studies of TRSB in several systems, emphasizing the study of the pseudogap state of high temperature superconductors and the inverse proximity effect in superconductor/ferromagnet proximity structures.
\end{abstract}

\maketitle

\section{Introduction}

In a system with strong, short-range repulsion between electrons, it is natural to find a superconducting order parameter that changes sign in different regions of the Brillouin zone, so that the average of the order parameter over {\bf k} is small or vanishing.  When the averaged order parameter over the entire Fermi surface of the material yields zero, that superconductor is deemed unconventional \cite{sigristueda}. Thus, non-$s$-wave superconductors, in which the phase of the order parameter changes sign, are unconventional. Although many of the earliest superconductors studied turned out to be unconventional, it was not until high-$T_c$ superconductors were discovered, and their order parameter shown to have $d$-wave symmetry, that the notion of unconventional superconductivity caught on as a special class.  $s$-wave superconductors inherently preserve time reversal symmetry; however, unconventional superconductors can be found in forms that break time reversal symmetry. In fact, we know now that the order parameter has $d_{x^2- y^2}$ symmetry in the cuprates, and likely in the ``115" heavy fermion superconductors and the ET  (bis(ethylenedithio)-tetrathiafulvalene, also known as BEDT-TTF) organics; $p$-wave symmetry in Sr$_2$RuO$_4$ (possibly $p_x \pm ip_y$), and probably the Bechgaard salts; and a complex pattern which is still not fully resolved in the newly discovered Fe-pnictides.  Often, this leads to the existence of gapless nodal quasiparticles in the superconducting state.  However, this is not necessary.  The presence of a secondary order parameter which nests the nodal points of the superconductor can open a gap in the quasiparticle spectrum without destroying superconductivity.  If time reversal symmetry is broken, so that the superconducting order parameter is complex, this typically insures a full gap in the quasiparticle spectrum.  While the additional types of order can produce various  side effects, the opening of a gap could, presumably, reduce the amount of low frequency dissipation in such a superconductor.

\section{Magneto-Optics as a Probe for Broken Time Reversal Symmetry}

Magneto-optical (MO) effects are described within quantum theory as the interaction of photons with electron spins through spin-orbit interaction (see e.g. \cite{pershan}). Macroscopically, linearly polarized light that interacts with magnetized media can exhibit both ellipticity and a rotation of the polarization state. The leading terms in any MO effects are proportional to the off-diagonal part of the ac conductivity: $\sigma_{xy}(\omega) = \sigma^{\prime}_{xy}(\omega) + i \sigma^{\prime \prime}_{xy}(\omega)$ \cite{pershan}. A finite MO effect measured in a material unambiguously points to time reversal symmetry breaking (TRSB) in that system. For example,  in an applied magnetic field, $\sigma_{xy}(\omega)$ is finite and proportional to the field. Its zero frequency limit is the known Hall coefficient of the material.

\begin{figure}[h]
\begin{center}
\includegraphics[width=1.0 \columnwidth]{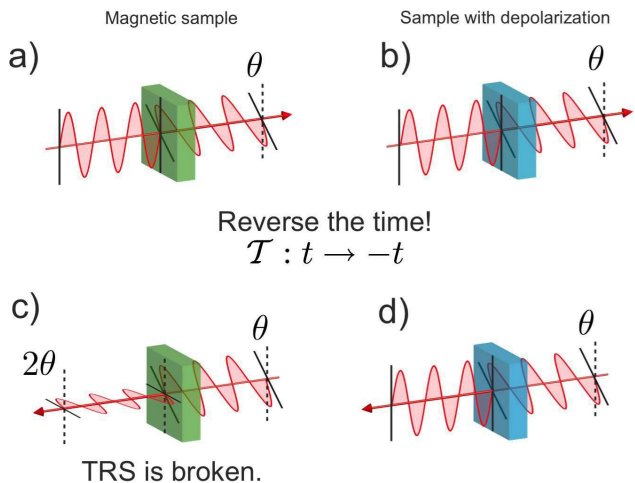}
\caption{ A simple test to distinguish a TRSB sample (a) from other reciprocal polarization-rotation effects (b). A mirror used in (a) will cause an additional rotation of polarization (c) while going through a non-TRSB sample will bring the polarization to its initial state (d).} 
\end{center}
\label{trsb}
\end{figure}

To demonstrate the relation between TRSB and MO effects, consider a transparent sample of thickness $\ell$, through which a linearly polarized plane wave of light (the electric field is ${\bf E}={\bf E}_0exp[i(kz - \omega t)]$) is transmitted.  If the material is circularly birefringent, i.e.~the complex index of refraction ($\tilde{n} = n+i\kappa$) for left and right circularly polarized light is different ($\tilde{n}_L \neq \tilde{n}_R$), the polarization will rotate by an angle \cite{argyres}

\begin{equation}
\theta_F= \frac{1}{2}\frac{\omega \ell}{c}\mathcal{R}e\left[\tilde{n}_L-\tilde{n}_R\right],
\end{equation}

\noindent where $\omega$ is the frequency of the light (in vacuum: $\lambda=2\pi c/\omega$).  This is called the Faraday effect and $\theta_F$ is the Faraday angle. The connection to TRSB arises when we want to distinguish this particular rotation of polarization from other, reciprocal effects that may also rotate the polarization. For example, a linearly birefringent material will also rotate the polarization when linearly polarized light that is not aligned with one of the principal axes of the material goes through it. To distinguish reciprocal effects from TRSB effects we apply a time-reversal operator, $\mathcal{T}$, to the light that emerges from our sample.  Since $\mathcal{T}$ is an antiunitary operator, it will transform ${\bf E}(z,t)$ to ${\bf E^*}(z,-t)$. This will result in a plane wave going in the opposite direction through the sample, an effect that can easily be simulated with a mirror as is depicted in Fig.~\ref{trsb}.

Indeed, placing a mirror after the sample and measuring the state of the polarization at the initial location after the beam has passed back through the sample has two possible results. If time-reversal symmetry is not broken, the light will go back to the initial state of linear polarization. However, if time-reversal symmetry is broken, i.e.~a true Faraday effect has occurred, the polarization at the origin after the beam has returned through the sample will read twice the Faraday angle.  This simple consideration not only demonstrates the connection between TRSB and magneto-optic effects (in this case the Faraday effect), but also suggests a method for measuring the amount of TRSB in a magnetically circular-birefringent material: one needs simply to compare two identical beams of light that counter-propagate through the test sample while traversing the exact same optical path.  The application of the above considerations to the polar Kerr effect (PKE), in which a rotation of polarization is detected for a beam of light reflected from a magnetic-circularly birefringent material, is straightforward and is given in ref.~\cite{argyres}.  The Kerr angle is given by

\begin{equation}
\theta_K=-\mathcal{I}m\left[ \frac{\tilde{n}_L - \tilde{n}_R}{\tilde{n}_L \tilde{n}_R-1}\right],
\label{kerr1}
\end{equation}

\noindent which in the case of weak absorption (i.e.~$n_L \gg \kappa_L$ and $n_R \gg \kappa_R$) reduces to

\begin{equation}
\theta_K \approx  \frac{4\pi}{n(n^2-1)\omega} \sigma^{\prime \prime}_{xy}(\omega).
\label{kerr2}
\end{equation}

We next discuss our unique method of testing for TRSB effects using a Sagnac interferometer, which relies exactly on the comparison between two counter-propagating beams through a magnetic-circularly birefringent material.

\section{The Sagnac Interferometer as a Magneto-Optic Sensor}

\begin{figure}[h]
\begin{center}
\includegraphics[width=1.0 \columnwidth]{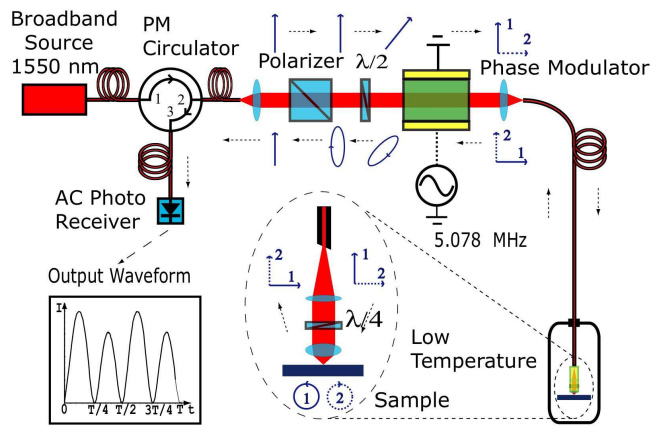}
\caption{ Experimental setup and the polarization states at various locations in the apparatus: vertical, in-plane polarization; horizontal, out-of-plane polarization; solid line, beam 1; dashed line, beam 2. } 
\end{center}
\label{setup}
\end{figure}

Our ability to perform high precision Faraday and polar Kerr effect measurements relies on the newly constructed zero-area-loop Sagnac interferometer. With this instrument we are able to probe non-reciprocal circular birefringence effects, while rejecting reciprocal effects to unprecedented accuracy.  Fig.~\ref{setup} shows a schematic of the design. The output of a very-short-coherence length ($\sim$30 $\mu$m) fiber-coupled superluminescent light emitting diode (SLED) centered at 1550 nm is routed by a fiber polarization-maintaining (PM) circulator to a Glan-Calcite polarizer, and becomes linearly polarized. The polarization is then rotated by a half-wave ($\lambda /2$) plate to a position that is 45$^o$  to the axes of a birefringent electro-optic modulator (EOM), which operates at $f_m = 5.078$ MHz. The length of the fiber strand is chosen to match twice the optical transit time of the system. After passing the EOM, the beam is split into two parts, with polarizations along the fast and slow axes of the EOM. They are then launched into the fast and slow axes respectively of a $\sim$10-m-long polarization maintaining (PM) fiber that is fed into a He-3 cryostat (base temperature $<$ 0.5 K). An aspheric lens focuses the light coming out of the fiber through a 100-$\mu$m-thick quartz quarter-wave ($\lambda /4$) plate into a spot with $1/e^2$ diameter in the range of $\sim$3 $\mu$m to $\sim$25 $\mu$m on the surface of the sample. The $\lambda /4$ plate is aligned at 45$^o$  to the axes of the PM fiber and converts the two orthogonally polarized beams into right- and left-circularly polarized light. The non-reciprocal phase shift $\phi_{nr}$ between the two circularly polarized beams upon reflection from the TRS-breaking sample is twice the Kerr rotation \cite{spielman1,spielman2,kdf} ($\phi_{nr} = 2 \theta_K$), while if a transparent sample is used with a mirror at its back, the non-reciprocal phase shift  between the two circularly polarized beams is four times the Faraday angle ($\phi_{nr} = 4 \theta_F$).  The same $\lambda /4$ plate converts the reflected beams back into linear polarizations, but with a net 90$^o$ rotation of polarization axis. In this way, the two beams effectively exchange their paths when they travel back through the PM fiber and the EOM to the polarizer. After passing the polarizer, the light is routed by the circulator to an AC-coupled photo-detector. Therefore the two beams travel precisely the same distance from source to detector, and should constructively interfere except for a small phase difference $\phi_{nr}$, which is solely from the TRS-breaking effect in the sample. Note that any light which did not follow the correct path (i.e.~due to back reflections and scattering, as well as polarization coupling due to birefringence of the sample or imperfections and misalignment of waveplates) will differ by many times the coherence length due to the birefringence of the PM fiber and EOM. This light may reach the detector, but can't interfere coherently with the main beams; it will at most add a constant background. Thus, the EOM serves as a convenient way to actively bias the interferometer to its maximum response and enable lock-in detection \cite{spielman1,spielman2}. The signal from the detector will contain even harmonics of $f_m$ proportional to the overall reflected intensity, and odd harmonics proportional to $\phi_{nr}$\cite{spielman1,spielman2}. Additional details of this apparatus, including its low temperature performance, are described in Xia {\it et al.} \cite{xia0}. We note that in some applications  we removed the focusing lens between the $\lambda /4$ plate and the sample in order to minimize the drift due to the Faraday effect in that lens. To summarize the current performance of the apparatus, we achieved a shot-noise limited sensitivity of 100 nanorad/$\sqrt{\rm Hz}$ at 10 $\mu$W of detected optical power from room temperature down to 0.5 K, with stability of better than 30 nanorad over a period of 50 hours, or better than 10 nanorad over a span of 8 hours if chosen properly during the day (the environment has strong influence on the ultimate performance of the electronics associated with the apparatus). 

\begin{figure}[h]
\begin{center}
\includegraphics[width=1.0 \columnwidth]{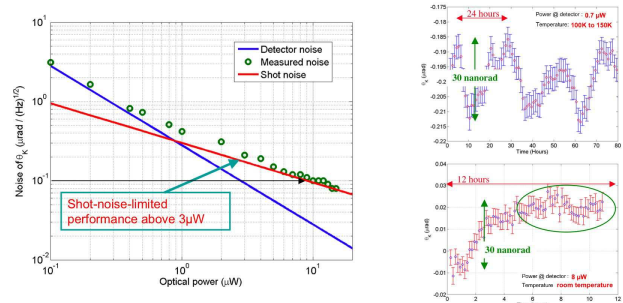}
\caption{ Noise figures and stability of the zero-area-loop Sagnac interferometer. a) Noise measurements of $\theta_K$ vs.~optical power. The reference lines of the detector noise and shot-noise indicate that the system becomes shot-noise limited below $\sim$ 3 $\mu$Watt. b) Stability of the system shows oscillations with a period of 24 hours bound by a drift of 30 nanorad (top), mostly occurring due to local air handling system. Note that in every 24 hours this effect is minimum over a long span of time (over 6 hours) in which the drift is bound by no more than $\sim$10 nanorad. } 
\end{center}
\label{perform}
\end{figure}

\section{Sr$_2$RuO$_4$}

Soon after the discovery of the layered-perovskite superconductor Sr$_2$RuO$_4$ \cite{maeno},  it was predicted to be have odd-parity pairing symmetry \cite{rice1,baskaran}. Subsequently, a large body of experimental results in support of odd-parity superconductivity has been obtained \cite{mackenzie}. The symmetry of the superconducting state is related simply to the relative orbital angular momentum of the electrons in each Cooper pair. Odd parity corresponds to odd orbital angular momentum and symmetric spin-triplet pairing. While {\it a priori} the angular momentum state can be  $p$ (i.e.~$L=1$), $f$ (i.e.~$L=3$), or even higher order \cite{annett,sigrist}, theoretical analyses of superconductivity in Sr$_2$RuO$_4$ favor the $p$-wave order parameter symmetry \cite{rice1,machida}.  There are many allowed $p$-wave states that satisfy the cylindrical Fermi surface for a tetragonal crystal, which is the case of Sr$_2$RuO$_4$ (see e.g.~table IV in \cite{mackenzie}). Some of these states break time-reversal symmetry (TRS), since the condensate has an overall magnetic moment due to either the spin or orbital (or both) parts of the pair wave function.  However, in their seminal paper, Rice and Sigrist \cite{rice1}, analyzing normal-state data as well, concluded that the most probable order parameter for Sr$_2$RuO$_4$ is {\bf d}({\bf p})=$\hat{\bf z}[p_x \pm ip_y]$, where following the convention of Balian and Werthamer \cite{balian} we used the three-component complex vector {\bf d(p)} to represent the superconducting gap matrix. 

In a $p$-wave superconductor each Cooper pair carries $\hbar$ amount of angular momentum and thus a  $p_x+ip_y$ superconductor would be viewed as an orbital ferromagnet with a uniform magnetization pointing perpendicular to the Ru-O planes.  While an ideal sample will not exhibit a net magnetic moment due to Meissner current screening,  surfaces and defects at which the Meissner screening of the TRS-breaking moment is not perfect and can result in a small magnetic signal \cite{sigrist}.  Indeed, muon spin relaxation ($\mu$SR) measurements on good quality single crystals of Sr$_2$RuO$_4$ showed excess relaxation that spontaneously appeared at the superconducting transition temperature.  The exponential nature of the increased relaxation suggested that its source is a broad distribution of internal fields, of strength $\sim$ 0.5 Oe, from a dilute array of sources \cite{luke1,luke2}. While TRS breaking is not the only explanation for the $\mu$SR observations, it was accepted as the most likely one \cite{mackenzie}.  At the same time, a $p_x+ip_y$ state is expected to result in edge currents that, while largely reduced due to opposing Meissner currents, are still expected to be measurable \cite{matsumoto,stone}. However, to date, neither Hall bar microscopy  \cite{bjornsson} nor scanning SQUID measurements \cite{kirtley} could observe the signature of such currents.  We note however that these two seemingly opposing results may still be reconciled if we accept the fact that the symmetry of the order parameter is not a single band, pure $p_x+ip_y$ one, and it may be further complicated by the presence of spin-orbit interaction. Since interpretation of both measurements is strongly model dependent, the conclusions implied by each of these experiments may not be conclusive. This possibility is not new, since the most anticipated effect for a pure $p_x+ip_y$ superconductor, namely the in-plane strong anisotropy of the upper critical field \cite{agterberg,mineev}, was never observed \cite{maeno1}.  The lack of large anisotropy for the low-field region of $H_{c2}$ between the (100) and (110) directions \cite{agterberg}, was suggested as evidence that more than one band participate in shaping the superconducting state \cite{mazin1,agterberg,mineev}. The need to look at all three bands of Sr$_2$RuO$_4$ has become even more imperative with the recent discovery that spin-orbit interaction in this system is as large as 100 meV in some parts of the Fermi surface \cite{haverkort,liu}. 

Irrespective of the details of the symmetry of the order parameter,  the issue of TRSB has to be resolved through an independent measurement.  As TRSB has considerable implications for the understanding of superconductivity in Sr$_2$RuO$_4$, establishing its existence, and in particular in the bulk, without relying on imperfections and defects is of utmost importance. The challenge is therefore to couple to the TRS-breaking part of the order parameter to demonstrate the effect unambiguously.

\begin{figure}[h]
\begin{center}
\includegraphics[width=1.0 \columnwidth]{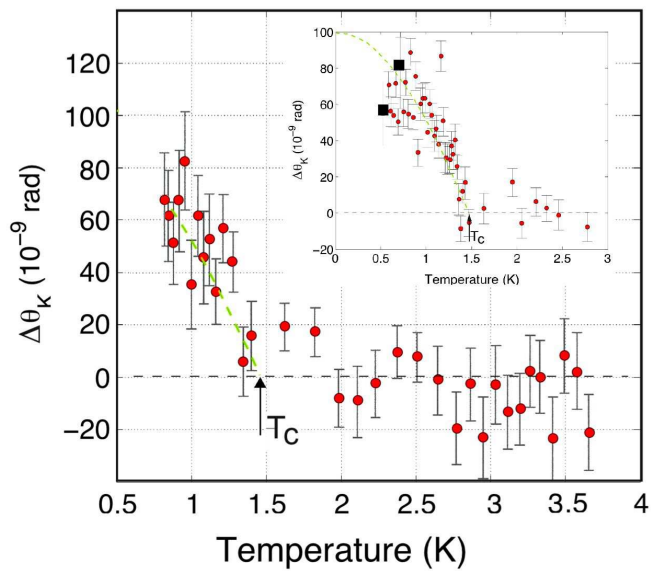}
\caption{ Representative results of PKE measurements of Sr$_2$RuO$_4$. Main panel shows the emergence of the PKE after zero-field cool.  Inset shows the an example of training the chirality with an applied field. A +93 Oe field cool was used, followed by a zero field warm-up (circles). The two solid squares represent the last two points just before the field was turned off. } 
\end{center}
\label{s2ro4}
\end{figure}

As explained above, measuring a finite polar Kerr effect (PKE) unambiguously points to TRSB.  However, early estimates of this circular dichroism in Sr$_2$RuO$_4$ suggested that this effect would be very small \cite{yipsauls}. To detect this very small PKE we used the zero-area-loop Sagnac interferometer at wavelength  $\lambda$=1550 nm described above. Our results show the emergence of a finite PKE at $T_c \approx 1.5$ K \cite{xia1}.  The size of the effect increases with decreasing temperature down to $0.5$ K, tracking the increase in magnitude of the superconducting order parameter.  Figure~\ref{s2ro4} has some of the important components of our results. First, the main panel shows the emergence of the PKE signal below $T_c$. While the signal above $T_c$ fluctuates around zero Kerr signal with no more than 20 nanorad, the increase below $T_c$ is roughly linear in $(T_c-T)$, starting to saturate at low temperatures and possibly pointing to a $\sim$100 nanorad signal at $T=0$, as seen in the inset. This linear dependence indicates a quadratic dependence on the order parameter, as expected from symmetry arguments \cite{yakovenko,mineev1,roy}.  We further note that zero-field cool measurements have shown a variety of results where the largest signal was the same as the field cool results. Moreover, the fact that a $\sim$100 Oe field cool result is the same as zero field and does not change whether the field is applied or removed (see inset of Fig.~\ref{s2ro4}) suggests that the effect is not a vortex effect but rather is intrinsic to the superconducting state.  Another important detail embedded in Fig.~\ref{s2ro4} is that the main panel was taken at incident power of $\sim$6 $\mu$Watt, while the data in the inset (as well as many other data of zero field cool or trained data at field cool) was taken at $\sim$0.7 $\mu$Watt. The fact that both data show the same size of effect shows that the PKE we observe is not a magneto-thermal effect. Based on the above results we conclude that TRS is broken in Sr$_2$RuO$_4$ below $T_c$. 

Our unambiguous results indicate that more theoretical work has to be done to reconcile all experimental results.  While it is widely accepted that strontium ruthenate is an unconventional superconductor and most likely exhibits triplet pairing,  puzzles remain in trying to connect this with the various measurements aimed at observing TRSB (see e.g. ref. \cite{kallin1}).  Concentrating on the Kerr effect, a pure $p_x+ip_y$ state with no disorder will result in zero signal (for a pure plane-wave-light) \cite{roy,lutchyn,ashby}. This consequence of gauge invariance corrects earlier finite results \cite{yakovenko,mineev1}. However, The materials parameters of all samples used were carefully measured, and it is clear that the mean free path and scattering time are finite. To take this into account, Goryo calculated the off-diagonal component of a current-current correlation function induced by impurity scattering in a chiral $p$-wave condensate \cite{goryo}. The skew-scattering-type diagrams give the leading contribution to a finite off-diagonal conductivity. The calculation of $\tilde{n}_L$ and $\tilde{n}_R$ with the actual material parameters were used to calculate $\theta_K$ according to Eqn.~\ref{kerr1}. The theoretical Kerr angle calculated by Goryo agrees to within 15$\%$ with our experimental result. These results were recently confirmed by Yakovenko and Lutchyn \cite{yakovenko1}. Since in this impurity-induced mechanism, the PKE would be suppressed or zero for any state other than the chiral $p$-wave state, these combined results give strong support for a TRSB chiral $p$-wave state in Sr$_2$RuO$_4$.

\section{YBa$_2$Cu$_3$O$_{6+x}$}

One of the greatest outstanding puzzles in modern condensed matter physics is high-temperature superconductivity (HTSC).  Understanding HTSC promises to show the way to a more general understanding of strongly correlated electron systems.  While there are many hallmark features associated with HTSC, the most prominent is the occurrence of a pseudogap for underdoped cuprates. This pseudogap state \cite{pseudoreview} is marked by the onset of anomalous behavior of many of the system's properties, including magnetic \cite{alloul}, transport \cite{ito}, thermodynamic \cite{loram}, and optical properties \cite{basov} below a temperature, $T^*$, large compared to the superconducting (SC) transition temperature, $T_c$. The origin of the pseudogap is a challenging issue, and it is believed to hold an important clue to our understanding of the mechanism behind HTSC \cite{pseudoreview}. Two major classes of theories have been introduced in an attempt to describe the pseudogap state.  In the first class, the pseudogap temperature $T^*$ represents a crossover into a state with preformed pairs with a $d$-wave gap character \cite{lee1,ek}. In the second class of theories, $T^*$ marks a true broken symmetry that ends at a quantum critical point, typically inside the superconducting dome. While the low-doping phase may compete with superconductivity, the disordered phase may provide fluctuations that are responsible for the enhanced transition temperature in the cuprates. Examples include competing phases of charge and spin density waves \cite{sketal}, or charge current loops which either do  \cite{sudip} or do not \cite{varma} break translational symmetry.  Concentrating on the pseudogap phase, some of the recent questions associated with its occurrence include i) whether the pseudogap is independent of superconductivity, ii) whether it is magnetic in origin, iii) whether it marks a broken symmetry state, and iv) whether it leads to a quantum critical point at some critical doping. 

Again, high resolution Kerr effect measurements add important information that, in conjunction with other data, can provide new insight into the nature of the pseudogap state. Indeed, our recent studies of PKE on YBa$_2$Cu$_3$O$_{6+x}$ suggest the presence of a small spontaneous static magnetic response intimately related to the pseudogap transition \cite{xia2}. Our most significant result is that we find a trend in the location and size of the Kerr signal that appears at the onset of pseudogap, and which seems to cross the superconducting dome, thus suggesting that the line defined by the pseudogap temperature $T^*$ ends at a quantum critical point near optimal doping. In addition, we find strong evidence that a novel magnetic state with broken time reversal symmetry already exists in the system well above room temperature. This state is coupled to the pseudogap order parameter, thus allowing us to observe the broken symmetry state at $T^*$ when the temperature is lowered.

High quality YBa$_2$Cu$_3$O$_{6+x}$ single crystals with $x=0.5$ (ortho-II, $T_c=59$K), $x=0.67$ (ortho-VIII, $T_c=65$K), $x=0.75$ (ortho-III, $T_c=75$K), and $x=0.92$ (ortho-I, $T_c=92$K)  \cite{liang}, as well as high quality thin films were used in this study. The crystals, in the form of ($ab$-plane) platelets several millimeters on a side and a fraction of a millimeter thick ($c$-direction), were mechanically detwinned. X-ray diffraction measurements indicate typical uninterrupted chain lengths of about 120$\times$b \cite{liang}.  

\begin{figure}[h]
\begin{center}
\includegraphics[width=1.0 \columnwidth]{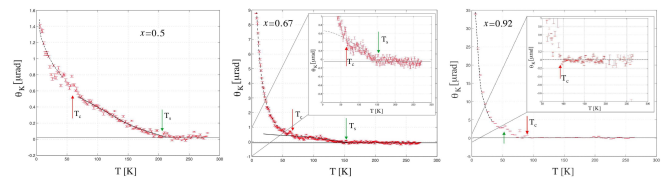}
\caption{ Kerr effect data taken on 3 crystals with different compositions. Insets for the left two panels we also show an enlarged region above $T_c$.} 
\end{center}
\label{highfield}
\end{figure}

Fig.~\ref{highfield} shows a typical cycle for three compositions.  Samples were first cooled in a 5 T field down to 4 K. At that temperature the field was removed and Kerr data was collected while warming the sample back to room temperature. Three regimes are clearly observed in the data. To best see them, let us first concentrate on the middle panel where we display the data for $x=0.67$. The low-temperature regime shows a large signal that decays exponentially when the temperature increases, until it almost disappears at $T_c=65$ K. We interpret the signal in this regime as due to vortices. We note that pinning forces are strong for this intermediate-anisotropy sample, and thus a relatively large remanent magnetization is expected when the field is removed at $T \sim T_c$. A small ($\sim$500 nanorad) remanent signal left at  $T_c$ decreases with increasing temperature until it completely disappears at a higher temperature $T_s$. Above $T_s$ we cannot detect any Kerr signal to a resolution of 50 nanorad. 

On the left in Fig.~\ref{highfield} we plot the result of Kerr data after a 5 T cooldown for a very underdoped sample ($x=0.5$). The relatively small Kerr enhancement below $T_c$ is in agreement with the large anisotropy, and therefore weak pinning, of this composition.  We note that $T_s$ for this composition is higher.  On the right in Fig.~\ref{highfield} we plot the result for an optimally-doped sample ($x=0.92$).  While we note the very strong Kerr signal at low temperatures, in agreement with the very strong pinning -- and hence the large remanent magnetization -- of this composition, we also note the fact that no remanent signal is detected above $T_c$ to a resolution of $\sim$50 nanorad. This is very different than for the underdoped samples.  An examination of the signal below $T_c$ indicates a ``bump" in the Kerr signal at around 50 K. We will examine this feature below when comparing the zero-field-cool results.

\begin{figure}[h]
\begin{center}
\includegraphics[width=1.0 \columnwidth]{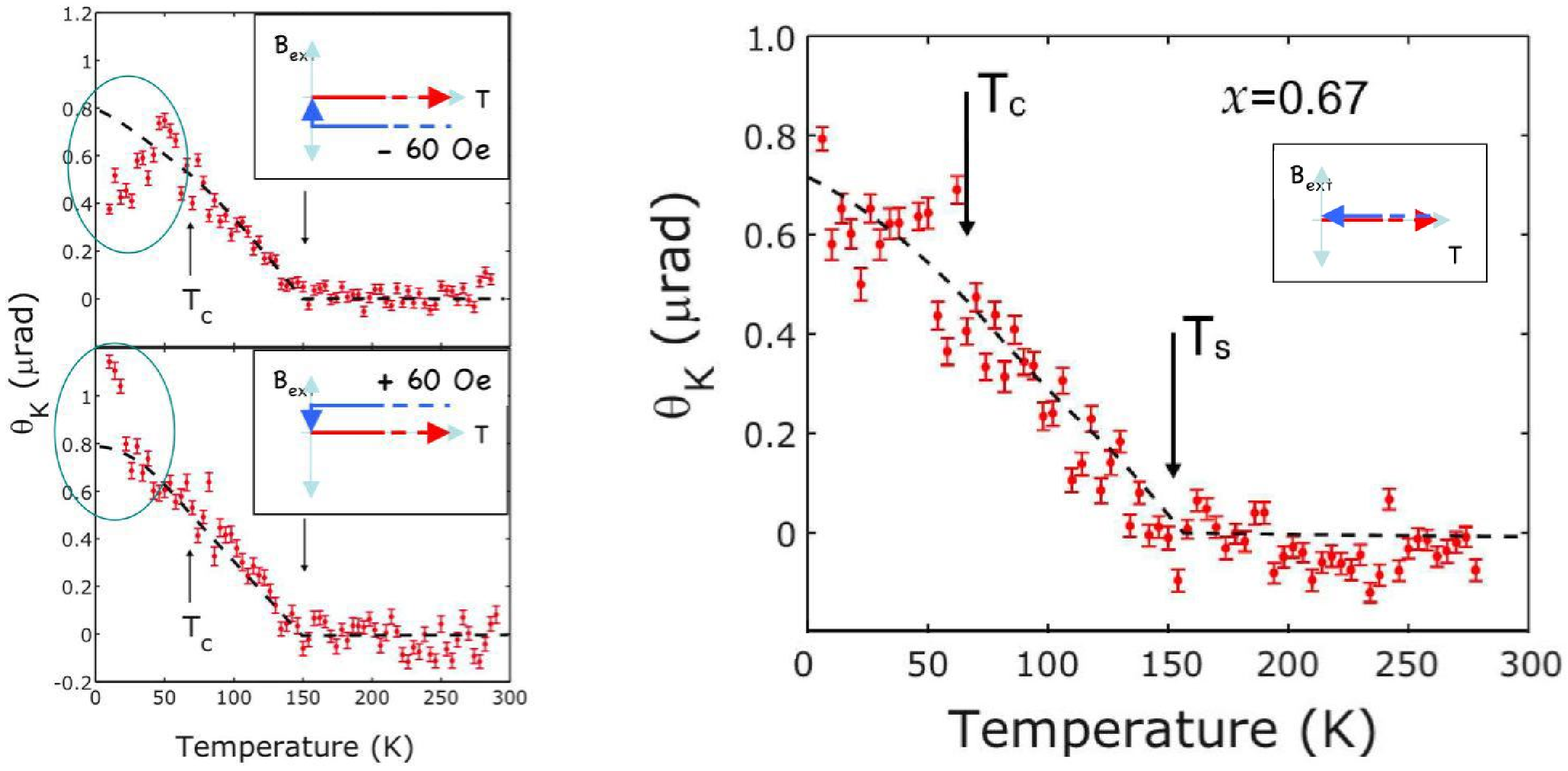}
\caption{ Polar Kerr effect of $x=0.67$ sample. Left: $\mp$60 Oe cooling-field data. Right: zero-field-cool data. } 
\end{center}
\label{ybco}
\end{figure}

Fig.~\ref{ybco} depicts the unusual magnetic properties  of the $x=0.67$ sample. Following the high-field cooldown (Fig.~\ref{highfield}), we examine the response of the system when cooled in small field. Here we used $\mp$60 Oe. First we observe that above $T_c$ the signal practically traces the signal measured after 5 T cooldown. However, while the sign of the signal above $T_c$ remains the same regardless of the direction of the cooling field, we note that below $T_c$ a small vortex-related signal does follow the magnetic field in which it was cooled down. On the right of Fig.~\ref{ybco} we show a true zero-field cool measurement (remanent field smaller than 3 milli-Oe). Again we trace the same signal above $T_c$, and this signal seems to continue smoothly through $T_c$, similarly to the average of the $\mp$60 Oe measurements.  This is a direct proof that a finite Kerr signal appears below $T_s$, and that this signal is an intrinsic effect in the material. The unusual inability to ``train" the sample with low fields above $T_c$ is found in all underdoped samples, with different large fields needed to flip the sign of the Kerr signal below $T_s$.  The nature of this unusual effect is currently under investigation.

\begin{figure}[h]
\begin{center}
\includegraphics[width=1.0 \columnwidth]{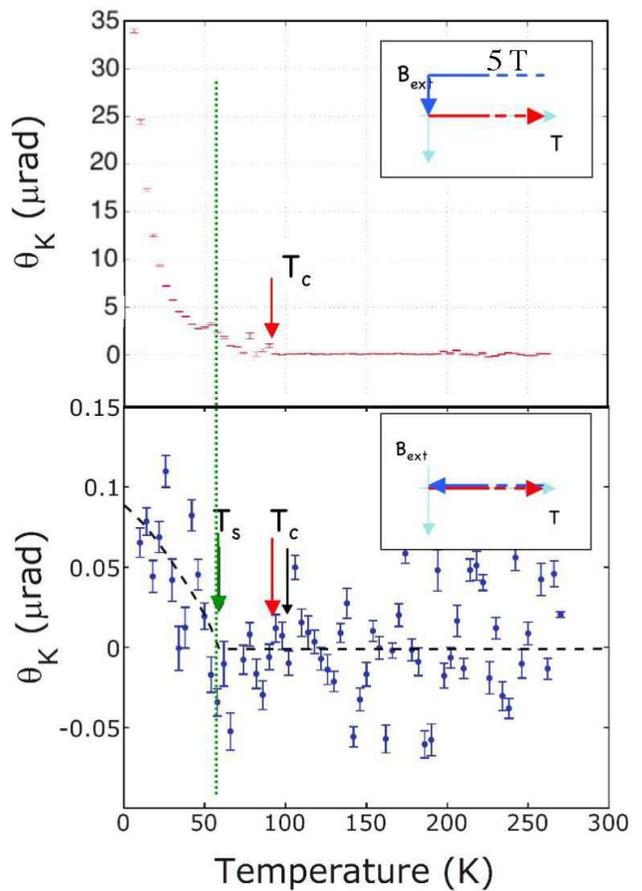}
\caption{ Zero-field measurements of Kerr effect on a $x=0.92$ sample. Top panel is a measurement after a 5 T cooldown while bottom panel  is a measurement after zero-field ($<$3 milli-Oe) cooldown. Note the correspondence between the "bump" in the Kerr measurement in the top panel and the onset of Kerr signal in the bottom panel.  } 
\end{center}
\label{optimal}
\end{figure}

Perhaps the most surprising result we have found, however, is in the study of the optimally-doped sample. Fig.~\ref{optimal} shows the high-field cool-down result of Fig.~\ref{highfield} together with the zero-field cooldown. First we note that in the zero-field cooldown there is no emerging signal above $T_c$ to within $\sim$50 nanorad, and that below $T_c$ there is a definite trend for the signal to be positive. While it is difficult to determine the actual temperature at which the Kerr signal appears, it is clearly in the vicinity of 50 K, and in fact coincides with the ``bump" found in the high-field cool that we mentioned earlier. We conclude that this finite Kerr signal appears for optimally doped samples below $T_c$.

\begin{figure}[h]
\begin{center}
\includegraphics[width=1.0 \columnwidth]{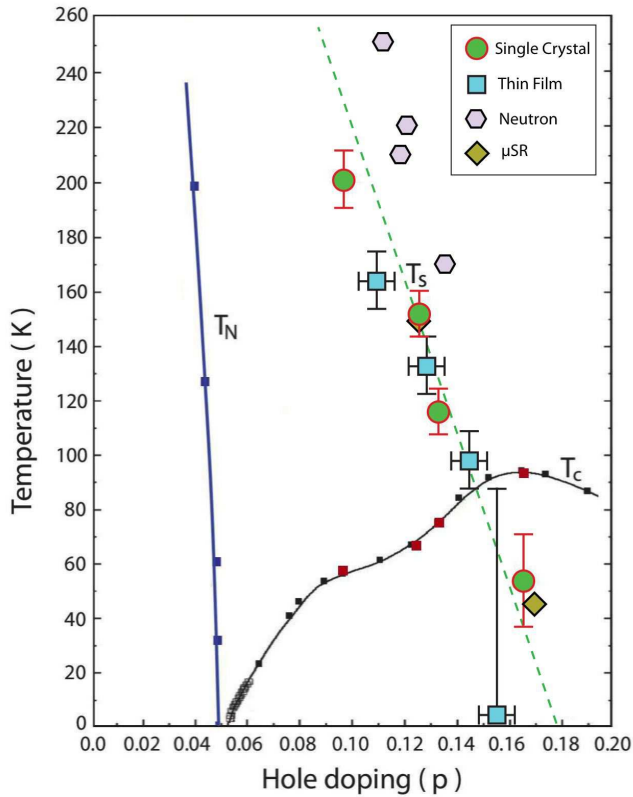}
\caption{ Compilation of the onset temperature of TRSB signal $T_s=T^*$ for the four single crystal samples of YBa$_2$Cu$_3$O$_{6+x}$ with $x=$0.5, 0.67, 0.75, and 0.92 (circles), as well as for four thin films samples (squares). The two diamonds represent the onset of $\mu$SR signal from Sonier {\it et al.} \cite{sonier}, and the four  hexagon symbols represent the neutron data of Fauqu\'{e} {\it et al.} \cite{fauque} and of Mook {\it et al.} \cite{mook}. Dashed line is a guide to the eye. $T_s$ for the near-optimally-doped thin film ($T_c =$91K) was difficult to determine and within 100 nanorad was consistent with zero (hence large error bar). Also shown are $T_c(p)$ (from \cite{liang}), and $T_N(p) $ (from  \cite{lavrov}). } 
\end{center}
\label{pd}
\end{figure}

To summarize our observations, we have identified a sharp phase transition at a temperature $T_s(x)$, below which there is a non-zero Kerr angle, indicating the existence of a phase with broken time reversal symmetry (TRS). The hole concentration dependence of $T_s$ is in close correspondence with the pseudo-gap crossover temperature, $T^*$, which has been identified in other physical quantities. In particular, as shown in Fig.~\ref{pd}, $T_s$ is substantially larger than the superconducting $T_c$ in underdoped materials, but drops rapidly with increasing hole concentration, so that it is smaller than $T_c$ in a near optimally doped crystal and extrapolates to zero at a putative quantum critical point under the superconducting dome. The magnitude of the Kerr rotation in YBa$_2$Cu$_3$O$_{6+x}$~is smaller by $\sim$4 orders of magnitude than that observed in other itinerant ferromagnetic oxides \cite{lsmomo,sromo}, and the temperature dependence is superlinear near $T_c$, suggesting that we are either not directly measuring the principal order parameter which characterizes the pseudo-gap phase in YBCO, or that we measure only its very small ``ferromagnetic-like" component. In addition we find a hysteretic memory effect that seemingly implies that TRS is broken in all cases at a still higher temperature (above room temperature), although no Kerr effect is detectable within our sensitivity at temperatures above $T_s$ \cite{xia2}.

The fact that $T_s$ marks the onset of a true symmetry breaking effect is lent support from the recent elastic neutron scattering measurements of Fauqu\'{e} {it et al.} \cite{fauque} and of Mook {\it et al.} \cite{mook}, as well as from earlier $\mu$SR measurements by Sonier {\it et al.} \cite{sonier}.  Using polarized elastic neutron diffraction, Fauqu\'{e} {\it et al.} identified a magnetic order in the YBa$_2$Cu$_3$O$_{6+x}$ system. This is done for each measured point by taking the difference between the neutron scattering intensity in the spin-flip channel, which measures the magnetic plus nuclear Bragg scattering, and the non-spin-flip channel, which measures the nuclear scattering only.  The new observed state was shown to have an unusual magnetic order in temperature and doping r\'{e}gimes that cover the range where the pseudogap occurs in YBa$_2$Cu$_3$O$_{6+x}$. In fact, the onset of the effect is  $\sim$ 30 K higher for similar dopings in the results from Fauqu\'{e} {it et al.}~as compared to our $T_s$, and is identical to our ortho-II results for $T_s$ in the case of Mook's result \cite{mook}.  The $\mu$SR results were taken from samples with dopings of $x=$0.67 and $x=$0.95, and for both samples, the onset temperature for increased muon relaxation matches our $T_s$. Moreover, the authors also note the existence of magnetic effects above the onset temperature that persist all the way to room temperature in both samples. Thus far, no idea has been proposed to explain this extra magnetic signal which (e.g.~in the case of neutron scattering) gives a background above which the new signal rides \cite{mook}. 

Fig.~\ref{pd} summarizes our results on crystals and films, as well as the neutron and $\mu$SR results.  From this curve it is very tempting to continue the straight line down to zero temperature, ending at a quantum critical point for $p \approx$ 0.18. 

To summarize this section, we reported in \cite{xia2} the discovery of a novel magnetic order across a wide range of doping in YBa$_2$Cu$_3$O$_{6+x}$. The new effect is ferromagnetic-like and onsets at a temperature that matches 
the pseudogap behavior in underdoped cuprates. We further find evidence that the line defined by the onset of this effect crosses the superconducting dome to appear below $T_c$ for a near-optimally-doped sample.  Finally, we find that this effect couples to another time reversal symmetry breaking effect that occurs at high temperatures and dictates the sign of the Kerr signal that appears at the pseudogap temperature.

\section{Superconductor/Ferromagnet Proximity Bilayers}

The proximity effect between a superconductor (S) and an itinerant ferromagnet (F) differs from the ordinary proximity effect (see e.g.~\cite{deutscher}) in the fact that, in itinerant ferromagnets, electrons with opposite spins belong to different bands that are shifted with respect to one another by the exchange interaction energy $J$ \cite{bergeret1,buzdin1,bulaevskii}. Thus, since a ferromagnet is inherently spin-polarized, singlet pairs from the superconductor will only penetrate a very short distance $L_M$ into the ferromagnet.  This penetration distance ranges from just few nanometers for strong ferromagnets such as Ni, Co, and Fe, to a few tens of nanometers for weaker ferromagnets such as NiCu alloys.  Some spectacular effects arise when a superconductor is sandwiched between two ferromagnets in a variety of S/F/S structures. For homogeneously magnetized ferromagnets, periodic $\pi$-phase shifts across the junction may occur as a function of the thickness of the ferromagnetic layer $d_F$, resulting in oscillatory behavior of the critical Josephson current $I_c(d_F)$  \cite{buzdin2,bergeret2,buzdin3}. This effect was observed experimentally in several S/F/S systems \cite{ryazanov,kontos,blum,robinson}. For non-homogeneously magnetized ferromagnets, the critical Josephson current was predicted to extend over larger distances, limited by the temperature length $L_T$ due to generation of an odd-triplet component  \cite{bergeret3,kadigrobov}.  In that case, the relevant spin projection  of the triplet component is $\pm$1, making the propagation of the superconducting order parameter into the ferromagnet possible, as it is now insensitive to the exchange field.  Experimental evidence for this effect was recently reported by Sosnin {\it et al.}  \cite{sosnin} and Keizer {\it et al.}~\cite{keizer}.

\begin{figure}[h]
\begin{center}
\includegraphics[width=1.0 \columnwidth]{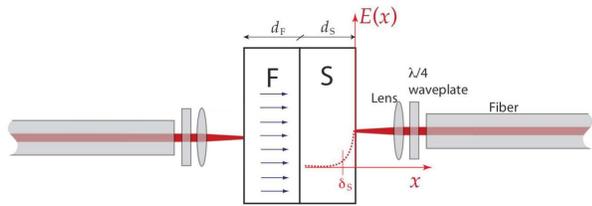}
\caption{ Cartoon of the measurement scheme. Here both orthogonal linear polarizations of light exiting the fiber \cite{xia0} become circularly polarized going through the  $\lambda$/4 plate and are then focused on the sample using a lens. The electric field of the incident light (represented with dashed line) penetrates a distance $\delta_S \ll d_S$ into the superconductor, thus is not sensitive to the spins in the F layer. \label{sch}} 
\end{center}
\end{figure}

While much of the work on S/F proximity effect focused on the penetration of the superconducting order parameter into the ferromagnet, very little was done to understand the penetration of the ferromagnetic order parameter, i.e.~the uniform magnetization, into the superconductor. For example, in the case of an induced triplet component, a novel proximity effect will result  from the zero spin projection \cite{bergeret4}. The theory in this case predicts an induced magnetization in the superconductor that can vary between states that either fully screen \cite{bergeret5}  or anti-screen \cite{bergeret4}  the magnetization of the ferromagnet, depending on the microscopic parameters of the system. The experimental observation of this so-called ``inverse proximity effect" has been viewed as a grand challenge in the field, as it would provide an important complementary confirmation of the possible triplet pairing in S/F structures. However, when designing an experiment to monitor the magnetization in the superconductor, one must bypass the large magnetic signal in the ferromagnet, which would otherwise tend to overwhelm any magnetic signal originating in the superconducting layer.

\begin{figure}[h]
\begin{center}
\includegraphics[width=1.0 \columnwidth]{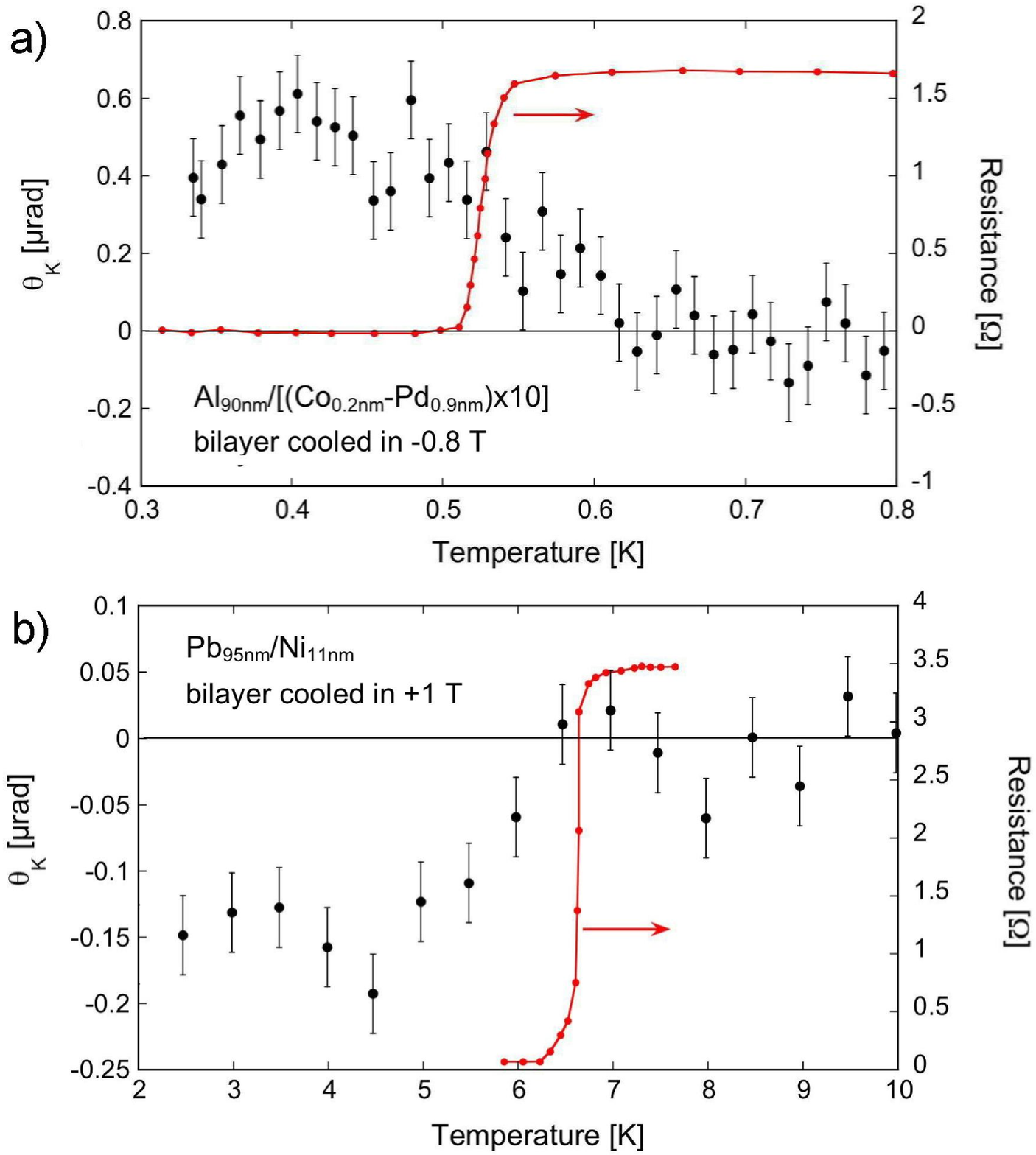}
\caption{ a)   Kerr effect measurement of a Al/(Co-Pd) bilayer system with 90 nm of Al. Sample was first cooled in a field of -0.8 T down to 10 K. The field was then turned down to zero, and the sample was cooled further to 0.3 K. Data was taken as the sample warmed up. For both samples we also show the resistive transition.  b) Kerr effect measurement of the Pb/Ni bilayer system. Sample was first cooled in a +1 T field down to 10 K. The field was then turned down to zero, and the sample was cooled further to 0.3 K. Data was taken as the sample warmed up. Also shown is the resistive transition. Note that in both cases the Kerr response indicates a magnetization that opposes the magnetization of the ferromagnetic layer.  \label{ipe}} 
\end{center}
\end{figure}

The zero-loop Sagnac interferometer is again the ideal instrument with which to address this challenge. Indeed, we recently reported direct experimental observations of the inverse proximity effect in Al/(Co-Pd) and Pb/Ni bilayers \cite{xia3}.  To show unambiguously that we detect a finite magnetization signal in the superconducting layer of a S/F bilayer structure, we performed magneto-optical Kerr effect measurements using light with an optical-penetration depth that is much smaller than the thickness of the superconducting layer, thus ensuring that the incident light does not interact with the moments in the ferromagnetic layer (see Fig.~\ref{sch}). We measure a finite signal that seems to onset below the superconducting transition temperature, $T_c$, of the bilayer and increases with decreasing temperature. For the Pb/Ni system, for which the superconducting coherence length $\xi_S$ is slightly smaller than the thickness of the superconducting film, the size of the effect is very small, of order 150 nanorad of optical rotation. For the Al/(Co-Pd) system the effect is much larger and increases with size as the temperature is lowered, in accordance with predictions by Bergeret {\it et al.} \cite{bergeret5}, as shown in Fig.~\ref{ipe}. To establish that we indeed measure the inverse proximity effect, we first note that the observed signal cannot result from a simple Meissner response to the magnetized ferromagnetic film. The observed signals are simply too large to result from screening currents. Moreover, because the films are thin, they can at most exist in a vortex state. However, vortices induced by the magnetized ferromagnetic films would result in a Kerr effect reflecting the magnetized ferromagnet, and thus would be opposite in sign to what we actually find experimentally. Note further that the magnetic field anywhere above the ferromagnetic film is vanishingly small due to the large aspect ratio, which results in a demagnetization factor very close to unity.

To summarize this section, we have observed the inverse proximity effect, in which the magnetization in a ferromagnetic film induces a magnetization that is much smaller and opposite in sign in the superconducting layer of a S/F bilayer proximity system. This observation may lead to a more quantitative description of the general S/F proximity effect.

\section{Conclusions and Outlook}

In this paper we have reviewed several recent measurements of TRSB in unconventional superconductors. Following a discussion of the novel Sagnac apparatus, we showed evidence for TRSB in the superconducting state of Sr$_2$RuO$_4$, we showed results that support a true phase transition at the pseudogap temperature in YBCO high-T$_c$ superconductors, and we introduced a scheme that allowed the first observation of the inverse proximity effect in superconductor/ferromagnet bilayers. We believe that these three different experiments, beyond the individual accomplishment for each of these studies, also demonstrate the power of the zero-loop Sagnac interferometer. 

At present we are able to reach a sensitivity of 100 nanorad working at a shot-noise limited power of 10 $\mu$Watt, allowing us to perform very low temperature measurements at the unprecedented sensitivity of a few nanorad.  However, the utilization of the instrument can be extended further with the addition of a scanning mechanism and a near field capability. While at present we are working at a diffraction limited beam of a few $\mu$m, it is feasible to add near field capabilities with at least $\lambda/$10 spatial resolution, as as previously shown by Dodge {\it et al.} \cite{dodge}.  The spatial resolution can be further improved with an instrument working at shorter wavelength. Indeed we are currently constructing such instrument operating at $\lambda =$830 nm. An instrument at this wavelength will also be especially well-suited for measurements of GaAs systems, since its wavelength coincides with the bandgap of this compound semiconductor.

Finally we note that the superconductor/ferromagnet proximity bilayer studies demonstrated the great selectivity of the Sagnac instrument, allowing us to probe the magnetization of the superconducting component only. We believe that this ability for selectivity will prove the instrument useful in additional applications.

\bigskip
\noindent {\bf Acknowledgements:}  Stimulating discussions with Steven Kivelson and Srinivas Raghu are greatly appreciated.   Initial fabrication of the Sagnac system was  supported by Stanford's Center for Probing the Nanoscale, NSF NSEC Grant 0425897. Work was supported by the Department of Energy Grant  DE-AC02-76SF00515. AP acknowledges support from the US-Israel BSF and the Israel Ministry of Science .


\begin{thebibliography}{99}

\bibitem{sigristueda}
M. Sigrist and K. Ueda, Rev. Mod. Phys. 63, 239 (1991).

\bibitem{pershan}
P.S. Pershan, J. Appl. Phys. 38, 1482 (1967).

\bibitem{argyres}
P.N. Argyres, Phys. Rev. 97, 334 (1955).

\bibitem{spielman1}
S. Spielman, K. Fesler, C.B. Eom, T. H. Geballe, M. M. Fejer, and A. Kapitulnik, Phys. Rev. Lett. 65, 123 (1990).

\bibitem{spielman2}
S. Spielman, J.S. Dodge, L.W. Lombardo, C.B. Eom, M.M. Fejer, T.H. Geballe, and A. Kapitulnik,  Phys. Rev. Lett.  68, 3472 (1992).

\bibitem{kdf}
A. Kapitulnik, J.S. Dodge, and M.M. Fejer, J. Appl. Phys. 75, 6872  (1994).

\bibitem{xia0}
Jing Xia, Maeno Yoshiteru, Peter T. Beyersdorf, M. M. Fejer, Aharon Kapitulnik, Phys. Rev. Lett. 97, 167002 (2006).

\bibitem{maeno}
Y. Maeno, H. Hashimoto, K. Yoshida, {\it et al.}, Nature  London  372, 532 (1994).

\bibitem{rice1}
T. M. Rice, M. Sigrist, J. Phys. Cond. Mat. 7, L643 (1995).

\bibitem{baskaran}
G. Baskaran, Physica B 223Ð224, 490 (1996).

\bibitem{mackenzie}
for a review see: Andrew Peter Mackenzie and Yoshiteru Maeno,  Rev. Mod. Phys. 75, 657 (2003). 

\bibitem{nelson}
K.D. Nelson, Z.Q. Mao, Y. Maeno, and Y. Liu, Science 306, 1151 (2004).

\bibitem{annett}
J.F. Annett, Adv. Phys. 39, 83 (1990).

\bibitem{sigrist}
M. Sigrist and K. Ueda, Rev. Mod. Phys. 63, 239 (1991).

\bibitem{machida}
K. Machida, M. Ozaki, and T. Ohmi, J. Phys. Soc. Jpn. 65, 3720 (1996).

\bibitem{balian}
R. Balian, and N. R. Werthamer,  Phys. Rev. 131, 1553 (1963).

\bibitem{luke1}
G.M.Luke, Y.Fudamoto, K.M.Kojima, M.I.Larkin, J.Merrin, B.Nachumi, Y.J.Uemura, Y.Maeno, Z.Q.Mao, Y.Mori, H.Nakamura, M.Sigrist, Nature 394, 558 (1998).

\bibitem{luke2}
G.M. Luke, Y. Fudamoto, K. M. Kojima, M. I. Larkinb, B. Nachumib, Y. J. Uemura, J. E. Sonierc, Y. Maenod, Z. Q. Maod, Y. Morid and D. F. Agterberge, Physica B 289, 373 (2000).

\bibitem{matsumoto}
M. Matsumoto and M. Sigrist, J. Phys. Soc. Jpn. 68, 994 (1999).

\bibitem{stone}
M. Stone and R. Roy, Phys. Rev. B 69, 184511 (2004).

\bibitem{bjornsson}
P.G. Bjornsson, Y. Maeno, M.E. Huber, and K.A. Moler, Phys. Rev. B 72, 012504 (2005).

\bibitem{kirtley}
J. R. Kirtley, C. Kallin, C. W. Hicks, E.-A. Kim,  Y. Liu,  K. A. Moler, Y. Maeno, and K. D. Nelson, Phys. Rev. B 76, 014526 (2007).

\bibitem{agterberg}
D.F. Agterberg, Phys. Rev. B 64, 2502 (2001).

\bibitem{mineev}
V.P. Mineev, Phys. Rev. B 77, 064509 (2008), and erratum in Phys. Rev B 77, 139901 (2008).

\bibitem{maeno1}
M. A. Tanatar, M. Suzuki, S. Nagai, Z. Q. Mao, Y. Maeno, and T. Ishiguro, Phys. Rev. Lett. 86, 2649 - 2652 (2001)

\bibitem{mazin1}
I. I. Mazin, D. A. Papaconstantopoulos, and D. J. Singh, Phys. Rev. B 61, 5223  (2000).

\bibitem{haverkort}
M.W. Haverkort, I.S. Elfimov, L.H. Tjeng, G.A. Sawatzky, A. Damascelli, Phys. Rev. Lett. 101, 026406 (2008).

\bibitem{liu}
Guo-Qiang Liu, V. N. Antonov, O. Jepsen, O.K. Andersen, Phys. Rev. Lett. 101, 026408 (2008).

\bibitem{yipsauls}
 S. K. Yip and J. A. Sauls, J. Low Temp. Phys. 86, 257 (1992).
 
\bibitem{xia1}
 Jing Xia, Yoshiteru Maeno, Peter Beyersdorf, M. M. Fejer, and A. Kapitulnik, Phys. Rev. Lett. 97, 167002 (2006).
 
\bibitem{yakovenko}
V.M. Yakovenko, Phys. Rev. Lett. 98, 087003 (2007); Roman M. Lutchyn, Pavel Nagornykh, Victor M. Yakovenko, Phys. Rev. B 77, 144516 (2008).

\bibitem{mineev1}
V.P. Mineev, Phys. Rev. B 76, 212501 (2007).

\bibitem{roy}
R. Roy and C. Kallin, Phys. Rev. B 77, 174513 (2008).

\bibitem{lutchyn}
Roman M. Lutchyn, Pavel Nagornykh, Victor M. Yakovenko, Phys. Rev. B 77, 144516 (2008).

\bibitem{ashby}
Phillip E. C. Ashby, Catherine Kallin, preprint, arXiv:0902.1704.

\bibitem{kallin1}
C. Kallin, A. J. Berlinsky, To appear in the proceedings of LT25 (Amsterdam, August 2008), arXiv:0902.2170.

\bibitem{goryo}
Jun Goryo, Phys. Rev. B 78, 060501(R) (2008).

\bibitem{yakovenko1}
V.M. Yakovenko and R.M. Lutchyn, private communication (2009).



\bibitem{pseudoreview}
For a recent review see e.g. M. R. Norman, D. P. Pines, and C. Kallin, Adv. Phys. 54, 715 (2005).

\bibitem{alloul}
H. Alloul {\it et al.}, Phys. Rev. Lett. 63, 1700 (1989); M. Takigawa, {\it et al.}, Phys. Rev. B 39, 7371 (1989); R. E. Walstedt, W. W. Warren Jr., Science 248, 1082 (1990).

\bibitem{ito}
T. Ito {\it et al.}, Phys. Rev. Lett. 70, 3995 (1993).

\bibitem{loram}
J.W. Loram {\it et al.}, Physica C 235-240, 134 (1994).

\bibitem{basov}
D.N. Basov and T. Timusk, Reviews of Modern Physics 77, 721 (2005).

\bibitem{lee1}
P. A. Lee, Physica C 317-318, 194 (1999).

\bibitem{ek}
V. J. Emery and S. A. Kivelson, Nature 374, 434 (1995).

\bibitem{sketal}
For a review see e.g. S.A.Kivelson, {\it et al.}, Rev. Mod. Phys. 75, 1201 (2003).

\bibitem{sudip}
S. Chakravarty, {\it et al.}, Phys. Rev. B 64, 094503 (2001).

\bibitem{varma}
C. M. Varma, Phys. Rev. B 55, 14 554 (1997); Phys. Rev. Lett. 83, 3538 (1999); Phys. Rev. B 73, 155113 (2006).

\bibitem{xia2}
J. Xia, E. Schemm, G. Deutscher,   S. A. Kivelson, D. A. Bonn, W. N. Hardy, R. Liang, W. Siemons, G. Koster, M. M. Fejer, and A. Kapitulnik, Phys. Rev. Lett 100, 127002 (2008)

\bibitem{liang}
Ruixing Liang, D.A. Bonn and W.N. Hardy, Physica C 336, 57 (2000).

\bibitem{lavrov}
A.N. Lavrov, Y. Ando, and K. Segawa, Physica C 341-348, 1555 (2000).

\bibitem{lsmomo}
S. Yamaguchi, Y. Okimoto, K. Ishibashi, and Y. Tokura, Phys. Rev. B 58, 6862  (1998).

\bibitem{sromo}
L. Klein, {\it et al.}, Appl. Phys. Lett. 66, 2427 (1995); G. Herranz, {\it et al.},  J. Appl. Phys. 97, 10M321 (2005).


\bibitem{fauque}
B. Fauqu\'{e}, Y. Sidis, V. Hinkov, S. Pailhes, C.T. Lin, X. Chaud, Ph. Bourges, Phys. Rev. Lett. 96, 197001 (2006).

\bibitem{mook}
H.A. Mook, Y. Sidis, B. Fauqu\'{e}, V. Bal\'{e}dent, P. Bourges, Phys. Rev. B 78, 020506 (2008).

\bibitem{sonier}
 J. E. Sonier J. H. Brewer, R. F. Kiefl, R. I. Miller, G. D. Morris, C. E. Stronach,  J. S. Gardner,5 S. R. Dunsiger, D. A. Bonn, W. N. Hardy, R. Liang, R. H. Heffner, Science 292, 1692 (2001).
 

\bibitem{deutscher}
G. Deutscher, P.G. de Gennes, Proximity effects, in ÒSuperconductivity,Ó  edited by R. D. Parks ( Marcel Dekker, New York); P.G. de Gennes Rev. Mod. Phys. 36, 225 (1964).

\bibitem{bergeret5}
F.S. Bergeret, A.F. Volkov, K.B. Efetov, Phys. Rev. B 69, 174 504 (2004).

\bibitem{bergeret1}
F.S. Bergeret, A.F. Volkov, K.B. Efetov, Rev. Mod. Phys. 77,1321 (2005). 
  
\bibitem{buzdin1}
  A. Buzdin, Rev. Mod. Phys. 77, 935 (2005). 
  
\bibitem{bulaevskii}
L. N. Bulaevskii, V. V. Kuzii, and A. A.Sobyanin, JETP Lett. 25, 290 (1977).

\bibitem{buzdin2}
A. I. Buzdin, L. N. Bulaevskii, and S.V. Panyukov, PisÕma Zh. Eksp. Teor. Fiz. 35, 147 (1982) [JETP Lett. 35, 178 (1982)]. 

\bibitem{bergeret2}
F. S. Bergeret, A. F.Volkov, and K. B. Efetov, Phys. Rev. B64, 134506 (2001).
 
\bibitem{buzdin3}
A. I. Buzdin, B. Bujicic, and  M. Yu. Kupriyanov, Sov. Phys. JETP 74, 124 (1992)
  
\bibitem{ryazanov}
V.V. Ryazanov {\it et al.}, Phys. Rev. Lett. 86, 2427 (2001).
    
\bibitem{kontos}
T. Kontos, M. Aprili, J. Lesueur, and X. Grison, Phys. Rev. Lett. 89, 137007 (2002).
      
\bibitem{blum}
Y. Blum, A. Tsukernik, M. Karpovski, and A. Palevski, Phys. Rev. Lett. 89, 187004 (2002).
        
\bibitem{robinson}
J.W.A. Robinson {\it et al.}, Phys. Rev. Lett. 97, 177003 (2006).

\bibitem{bergeret3}
F.S. Bergeret, A.F. Volkov, K.B. Efetov, Phys. Rev.Lett. 78, 4096 (2001).

\bibitem{kadigrobov}
A. Kadigrobov, R.I. Shekhter, M. Jonson, Eur. Phys. Lett. 54, 394 (2001).

\bibitem{sosnin}
I. Sosnin, H. Cho, V.T. Petrashov, A.F. Volkov, Phys. Rev. Lett. 96,157 002 (2006).

\bibitem{keizer}
R.S. Keizer {\it et al.},  Nature 439, 825 (2006).

\bibitem{bergeret4}
F.S. Bergeret, A.F. Volkov, K.B. Efetov, Eur. Phys. Lett. 66, 111 (2004).

 \bibitem{xia3}
 Jing Xia, V. Shelukhin, M. Karpovski, A. Kapitulnik, and A. Palevski, Phys. Rev. Lett 102, 087004 (2009).
 
\bibitem{dodge}
J.S. Dodge, Ph.D. Thesis [Title: Sagnac Magneto-Optic Interferometry and the Magneto-Optics of Transition Metal Oxides], Stanford University (1997), unpublished.



\end{thebibliography}
\end{document}